\documentclass[num-refs]{wiley-article}
\usepackage{nth}

\usepackage{enumitem}
\newcommand\litem[1]{\item{\bfseries#1.\space}}
\usepackage{CJKutf8}
\usepackage[utf8]{inputenc}
\usepackage{wasysym}
\usepackage{algorithm}
\usepackage[noend]{algpseudocode}

\algnewcommand\algorithmicswitch{\textbf{switch}}
\algnewcommand\algorithmiccase{\textbf{case}}
\algnewcommand\algorithmicassert{\texttt{assert}}
\algnewcommand\Assert[1]{\State \algorithmicassert(#1)}%
\algdef{SE}[SWITCH]{Switch}{EndSwitch}[1]{\algorithmicswitch\ #1\ \algorithmicdo}{\algorithmicend\ \algorithmicswitch}%
\algdef{SE}[CASE]{Case}{EndCase}[1]{\algorithmiccase\ #1 \algorithmicdo}{\algorithmicend\ \algorithmiccase}%
\algtext*{EndSwitch}%
\algtext*{EndCase}%

\algnewcommand{\IIf}[1]{\State\algorithmicif\ #1\ \algorithmicthen}
\algnewcommand{\EndIIf}{\unskip\ \algorithmicend\ \algorithmicif}
\algnewcommand{\EElse}[1]{\State\algorithmicelse\ #1\ }
\algnewcommand{\EndEElse}{\unskip\ \algorithmicend\ \algorithmicelse}
\algnewcommand{\CCase}[1]{\State\algorithmiccase\ #1}
\algnewcommand{\EndCCase}{\unskip\ \algorithmicend\ \algorithmiccase}

\newcommand*\ErrorState{\texttt{error}}
\newcommand*\ValidState{\texttt{valid}}
\newcommand*\Lookup{\texttt{lookup}}

\usepackage{algorithmicx}
\usepackage{tikzsymbols}
\usepackage[binary-units]{siunitx}
\usepackage{booktabs}
\usepackage{subfloat,subfig}
\usepackage{todonotes}
\usepackage{color}
\usepackage{multirow}
\usepackage{enumitem}
\definecolor{lightgray}{rgb}{.9,.9,.9}
\definecolor{darkgray}{rgb}{.4,.4,.4}
\definecolor{purple}{rgb}{0.65, 0.12, 0.82}
\usepackage{arydshln}
\usepackage{etoolbox}
\usepackage{xcolor}
\usepackage{listings}
\lstdefinestyle{customc}{%
  belowcaptionskip=1\baselineskip,
  breaklines=true,
  xleftmargin=\parindent,
  language=C,
  showstringspaces=false,
  basicstyle=\small\ttfamily,
  keywordstyle=\bfseries\color{green!40!black},
  numberstyle=\tiny,
  commentstyle=\itshape\color{purple!40!black},
  identifierstyle=\bfseries\color{black},
  stringstyle=\color{orange},
   morekeywords={uint64_t,uint32_t,__m256i,__m128i,simd8,uint8_t,UINT64_C},
}

\newcommand*{\codepointu}[1]{\texttt{U+#1}}
\newcommand*{\codepointrange}[2]{\texttt{U+#1\ldots#2}}
\newcommand*{\utfbinary}[2]{\texttt{\begingroup\color{darkgray}#1$\vert$\endgroup#2}}
\newcommand*{\utfbinaryrange}[3]{\texttt{\begingroup\color{darkgray}#1$\vert$\endgroup#2\ldots#3}}
\newcommand*{\utfzerob}[2]{\ensuremath{\begingroup\color{darkgray}\mathtt{0b#1\vert}\endgroup#2}}
\newcommand*{\utfzerobrange}[3]{\ensuremath{\begingroup\color{darkgray}\mathtt{0b#1\vert}\endgroup\{ #2\ldots#3 \}}}
\newcommand*{\restartrowcolors}{%
  \ifhmode\unskip\fi
  \vadjust{%
    \global\rownum=0 %
  }%
}

\papertype{Research Article}

\title{Validating UTF-8 In Less Than One Instruction Per Byte}

\author[1\authfn{1}]{John Keiser}
\author[2\authfn{1}]{Daniel Lemire}

\affil[1]{Microsoft, Redmond, WA, 98052, USA}
\affil[2]{DOT-Lab Research Center, Universit\'e du Qu\'ebec (TELUQ), Montreal, Quebec, H2S 3L5, Canada}

\corraddress{Daniel Lemire, DOT-Lab Research Center, Universit\'e du Qu\'ebec (TELUQ), Montreal, Quebec, H2S 3L5, Canada}
\corremail{lemire@gmail.com}

\fundinginfo{Natural Sciences and Engineering Research Council of Canada, Grant Number: RGPIN-2017-03910}

\runningauthor{John Keiser and Daniel Lemire}

\begin{document}

\maketitle

\begin{abstract}
The majority of text is stored in UTF-8, which must be validated on ingestion. We present the \Lookup{} algorithm, which outperforms UTF-8 validation routines used in many libraries and languages by more than 10~times using commonly available SIMD instructions. To ensure reproducibility, our work is freely available as open source software.

\end{abstract}

\section{Introduction}


Unicode is the ubiquitous standard for text representation in software. It assigns a \textit{code point} (a number from 0 to \num{1114112}) to almost every character in every language, as well as formatting and symbols like whitespace characters and emojis.
UTF-8, in turn, is the  dominant format used to \emph{encode} Unicode text---to store or send it in a series of bytes via memory, disk or network~\cite{rfc3629}. For example, UTF-8 is in widespread use in XML and JSON documents, as well as in database systems like MySQL\@. Even more fundamentally, many recently introduced programming languages represent strings as UTF-8 by default (e.g., Rust, Go) while  established  languages have migrated to UTF-8 (Swift, Ruby). UTF-8 is more concise than other alternative Unicode formats such as UTF-16 and UTF-32.

All of these systems have to \textit{validate}  UTF-8 on ingestion. Invalid UTF-8 strings can cause various functions such as search or sort to fail; they may cause display problems in applications or web sites. More critically, invalid UTF-8 is a security risk~\cite{utfsecure}: e.g.,  Microsoft's web server (IIS) failed to validate the UTF-8 string used as URI which allowed attackers to access otherwise forbidden paths.  Whenever a database system or software program receives bytes that are meant to be UTF-8, they run a validation function. 

Validation is not a straightforward problem. UTF-8 uses between 1 and 4~bytes to encode each character, and there are many distinct error cases to check.
In our experience, most systems validate UTF-8 using relatively complicated sequences of branches. The speed of a branch-based approach depends on the input. We can exceed speeds of \SI{2}{\gibi\byte\per\second}, going as fast as \SI{4}{\gibi\byte\per\second} on ASCII content. Though such speeds may seem satisfactory, recent disks can sustain higher throughput (e.g., \SI{5}{\gibi\byte\per\second}) with networking speeds being even higher.  Generic compression libraries such as LZ4 can decompress text data at  \SI{5}{\gibi\byte\per\second}~\cite{lz4}. An engineer behind the high-performance ScyllaDB database system~\cite{suneja2019scylladb} concluded that \emph{UTF-8 validation can become a bottleneck under
heavy loads}~\cite{scyllaquote}.

\paragraph{Going Faster}
Starting with the Pentium~4 launched at the beginning of the century, commodity processors have acquired single-instruction-multiple-data (SIMD) instructions capable of working on wide registers (e.g., 128-bit, 256-bit or even 512-bit). These SIMD instructions have become ubiquitous,  being available in nearly all mobile processors and in all x64 processors. These instructions enable an efficient form of single-core parallelism that comes in addition to multi-core and memory-level parallelism~\cite{6495875}. 

There are many different ways to benefit from these SIMD instructions. 
Optimizing compilers often try to rewrite tight loops so that they use SIMD instructions, a process called autovectorization~\cite{nuzman2006auto}. Though autovectorization is a powerful approach, the compilers  often fail to autovectorize complex routines. Furthermore, compilers cannot produce compiled code that deviates from the semantics of the original source code. 
The programmer may also rely on libraries that were written with SIMD instructions in mind. Finally, a programmer may design algorithms specifically for SIMD instructions. Though such an approach requires in-depth knowledge of the available instructions and of their performance, our experience is that it provides the best performance, at the cost of greater development time.

Our main contribution is a novel SIMD-based algorithm to validate UTF-8 bytes at high speed. We consistently exceed \SI{10}{\gibi\byte\per\second} on x64 processors. To achieve these good results, we have extended an existing technique, vectorized classification, to do most of the validation using few instructions.

\section{UTF-8}

UTF-8 encodes a sequence of Unicode characters into variable-length sequences of bytes. 
We  use the word ``character'' as defined by the Unicode standard: a single character from the \emph{Universal Character Set}, which has been assigned a single code point. However, this convention does not always correspond to a single letter in a word, or a single visible ``glyph.'' Not only are some Unicode characters invisible (e.g., new-line and control characters), glyphs are sometimes formed by combining \textit{multiple} Unicode characters into a \texttt{grapheme}. The distinction is irrelevant to UTF-8 validation. 


\paragraph{Variable-Length Characters}
UTF-8 achieves complete ASCII backward compatibility by encoding ASCII characters  (\codepointrange{00}{7F}) as-is. Further, it ensures that all non-ASCII bytes have a high order bit of 1, so ASCII characters can always be identified by a most significant bit of 0. The character ``9'' is \texttt{00111001} in both ASCII and UTF-8.

Non-ASCII characters (\codepointrange{000080}{10FFFF}) start with a \emph{leading byte} indicating whether the character is encoded using two, three, or four bytes. It denotes this character length with the number of \emph{header bits}---where the most significant bits are a series of 1's followed by a 0.\footnote{We adopt the convention that ASCII bytes are leading bytes with no header bits.} Thus, \utfbinary{110}{00010} is the leading byte of a two-byte character, \utfbinary{1110}{1001} starts a three-byte character, and you may expect three more bytes after \utfbinary{11110}{000}.

The 1--3~remaining bytes of a multi-byte character are called \texttt{continuation bytes}, and have exactly one header bit. The choice to use up the two most significant bits with \texttt{10} was a deliberate tradeoff, preserving ASCII compatibility at the cost of space: because their most significant bit is 1, continuation bytes are never  mistaken for ASCII\@. The four-byte character ``\Smiley[][yellow]'' has leading byte \utfbinary{11110}{000} followed by three continuation bytes: \utfbinary{10}{011111}, \utfbinary{10}{011000} and \utfbinary{10}{000000}.

UTF-8 tries to concisely represent as many frequently used languages as possible in as few bytes as possible.  Two-byte characters (up to \codepointu{07FF}) can represent most Latin alphabets, and other alphabets like Hebrew and Arabic. Most characters in natural languages (including Chinese and Japanese) fit into at most 3~bytes (up to \codepointu{FFFF}). Unicode uses 4-byte characters to represent ``supplementals'' such as emojis.

\paragraph{Encoding the Value}
The character value itself is stored by disassembling it bitwise and inserting its bits into the unused (non-header) bits of the byte sequence. The bits are inserted in reverse (``big-endian'') order, with the most significant bits in the leading byte, and the lowest bits in the last continuation byte.
See Table~\ref{tab:utfexample}.
Decoding UTF-8 is just reassembling the character value, validating that sequences are well-formed.
Thus UTF-8 is independent from the \emph{endianness} of the processor and system. It is still allowed to prefix the string with a byte-order-mask (the byte sequence 0xEF,0xBB,0xBF) but it does not add any difficulty to the validation since this sequence is valid UTF-8~\cite{xia2013software}.

\begin{table}\centering
    \caption{Valid Unicode characters and corresponding UTF-8 ranges~\cite{rfc3629}}
    \label{tab:unicode}
    \begin{tabular}{lrlll}
        \toprule
        UTF-8 Bytes &Bits & Description   & Code Point        & UTF-8  \\
        \midrule
        1 & 7  & ASCII         & \codepointu{0000} & \texttt{00000000} \\
          &    &               & \codepointu{007F} & \texttt{01111111} \\
        \midrule
        2 & 11 & Latin         & \codepointu{0080} & \utfbinary{110}{00010} \utfbinary{10}{000000} \\
          &    &               & \codepointu{07FF} & \utfbinary{110}{11111} \utfbinary{10}{111111} \\
        \midrule
        3 & 16 & Asiatic       & \codepointu{0800} & \utfbinary{1110}{0000} \utfbinary{10}{100000} \utfbinary{10}{000000} \\
          &    &               & \codepointu{D7FF} & \utfbinary{1110}{1101} \utfbinary{10}{011111} \utfbinary{10}{111111} \\
        \midrule
          &    &               & \codepointu{E000} & \utfbinary{1110}{1110} \utfbinary{10}{000000} \utfbinary{10}{000000} \\
          &    &               & \codepointu{FFFF} & \utfbinary{1110}{1111} \utfbinary{10}{111111} \utfbinary{10}{111111} \\
        \midrule
        4 & 21 & Supplementary & \codepointu{010000} & \utfbinary{11110}{000} \utfbinary{10}{010000} \utfbinary{10}{000000} \utfbinary{10}{000000} \\
          &    &               & \codepointu{10FFFF} & \utfbinary{11110}{100} \utfbinary{10}{001111} \utfbinary{10}{111111} \utfbinary{10}{111111} \\
        \bottomrule
    \end{tabular}
 
\end{table}

\begin{table}\centering
    \caption{UTF-8 Character Examples: ASCII, two, three and four-byte characters}
    \label{tab:utfexample}
\begin{tabular}{lrrrr}\toprule
Label & Byte 1 & Byte 2 &Byte 3  &Byte 4 \\\midrule
Text & 9 \small{(\codepointu{0039})} & & & \\
Binary & \texttt{00111001} & & & \\
UTF-8 & \texttt{00111001} & & & \\
\hline
Text & \cent{} \small{(\codepointu{00A3})} & & & \\
Binary & \texttt{10} & \texttt{100011} & & \\
UTF-8 & \utfbinary{110}{00010} & \utfbinary{10}{100011} & & \\
\hline
Text & \begin{CJK}{UTF8}{bsmi}鏡\end{CJK} \small{(\codepointu{93E1})} & & & \\
Binary & \texttt{1001} & \texttt{001111} & \texttt{100001} & \\
UTF-8 & \utfbinary{1110}{1001} & \utfbinary{10}{001111} & \utfbinary{10}{100001} & \\
\hline
Text & \Smiley[][yellow] \small{(\codepointu{1F600})} & & & \\
Binary & \texttt{00} & \texttt{011111} & \texttt{011000} & \texttt{000000} \\
UTF-8 & \utfbinary{11110}{000} & \utfbinary{10}{011111} & \utfbinary{10}{011000} & \utfbinary{10}{000000} \\
\bottomrule
\end{tabular}
\end{table}

\paragraph{UTF-8 Sortability}

UTF-8 is \textit{normalized}: there is only one way to write a Unicode string in UTF-8. Because of this byte-for-byte stability, UTF-8 strings are byte-sortable and byte-comparable. Two strings form the same sequence of characters if and only if their bytes are all the same. A string is likewise considered larger than another in Unicode lexicographical order if its first non-equal byte is larger.

This helps with compatibility: existing libraries like hash tables, and programs like \texttt{grep}, can be easily adapted to UTF-8, and often work without modification. It is also a security feature. There is only one way to represent a given character (such as the null character or the \texttt{/} character) and so validating the content of strings for security is easier.\footnote{A single \emph{visual} character, or glyph, may be represented by more than one \emph{sequence} of Unicode characters. This is not relevant to UTF-8, which operates at the Unicode character level.}

\section{Validating UTF-8}
\label{sec:validatingutf8}

A validator must step through each character in a UTF-8 document, checking for violations of each of these rules:

\begin{enumerate}[label=\emph{\alph*)}]
\litem{5+ Byte} The leading byte must have fewer than 5~header bits. 
\litem{Too Short} The leading byte must be followed by N-1 continuation bytes, where N is the UTF-8 character length. 
\litem{Too Long} The leading byte must not be a continuation byte. 
\litem{Overlong} The decoded character must be above \codepointu{7F} for two-byte characters, \codepointu{7FF} for three-byte characters, and \codepointu{FFFF} for four-byte characters. 
\litem{Too Large} The decoded character must be less than or equal to \codepointu{10FFFF}. 
\litem{Surrogate} The decoded character must be not be in \codepointrange{D800}{DFFF}. 
\end{enumerate}
The rules can be usefully separated into three kinds: malformed byte sequences, invalid Unicode characters, and overlong byte sequences.

\paragraph{Malformed Byte Sequences}
Any UTF-8 character must either be an ASCII byte, or a byte with 2--4~header bits followed by 1--3~continuation bytes---no more, and no less.
The easiest such error to detect is 5~or more header bits. These include \utfbinary{111110}{}, \utfbinary{1111110}{}, \utfbinary{11111110}{} and \utfbinary{11111111}{}. Out-of-order sequences and sequences with the wrong number of bytes are also invalid.  See Table~\ref{table:malformedbytesequenceexamples}.

\begin{table}
\caption{Examples of Malformed Byte Sequences}
\label{table:malformedbytesequenceexamples}
\begin{tabular}{llp{5cm}}\toprule 
type & byte sequence & \\\midrule
Too Long & \texttt{00111001} \utfbinary{10}{000000} & The continuation byte is a ``stray'', that is not a part of any character. \\
Too Short & \utfbinary{1110}{1001} \utfbinary{10}{001111} \texttt{00111001} & There are only 2~bytes in a 3-byte character. \\
5-Byte & \utfbinary{111110}{10} \utfbinary{10}{010000} \utfbinary{10}{010000} \utfbinary{10}{000000} \utfbinary{10}{000000} & 5-byte character sequences are disallowed. \\\bottomrule
\end{tabular}
\end{table}

\paragraph{Invalid Unicode Characters}
\label{sec:invalidunicodecharacters}
A well-formed byte sequence can always be decoded into a code point, but even then, some code points represent \emph{invalid} Unicode characters. For example, Unicode only supports characters from \codepointrange{000000}{10FFFF}. Anything outside that range is \emph{too large} and therefore invalid. Since 4-byte characters can encode anything up to \codepointu{1FFFFF}, characters from \codepointrange{110000}{1FFFFF} are too large.

Additionally, UTF-8 disallows Unicode \textit{surrogate} characters (\codepointrange{D800}{DFFF}), which were designed to encode values larger than 16~bits in UTF-16. UTF-8 disallows these because it already has a way to encode characters larger than 16~bits, and surrogate support would break the normalization rule that there is only one way to encode a given code point.
See Table~\ref{table:invalidunicodecharacterexamples}.

\begin{table}
\caption{Invalid Unicode Character Examples}
\label{table:invalidunicodecharacterexamples}
\begin{tabular}{lll} \toprule
type & byte sequence & \\ \midrule
Surrogate & \utfbinary{1110}{1101} \utfbinary{10}{111000} \utfbinary{10}{000000} & \codepointu{D83D} \codepointu{DE00} is the surrogate pair for ``\Smiley[][yellow]'' (\codepointu{1F600}). \\
Too Large & \utfbinary{11110}{100} \utfbinary{10}{010000} \utfbinary{10}{000000} \utfbinary{10}{000000} & \codepointu{110000} is larger than the largest Unicode character.
\end{tabular}
\end{table}

\paragraph{Overlong Byte Sequences}
\label{sec:overlongbytesequences}

UTF-8 mandates that each character be encoded in the smallest number of bytes possible. Larger sequences would be well-formed, and represent valid Unicode characters, but they break the normalization rule.

Overlong byte sequences are violations of this rule. For example, the character ``a'' (\codepointu{61}) in a 3-byte character, padding it with zeroes: \utfbinary{1110}{0000}\utfbinary{10}{000001} \utfbinary{10}{100001}.  This is the only category of invalid UTF-8 that can occur even when the byte sequence is well-formed and represents a valid Unicode character.

\section{Branchy Range Validator}

A \emph{Branchy Range Validator} validates without decoding, walking the input character by character and checking that each byte in the character is in a specific range. It branches on the value of the first byte of each character, using it to decide how many continuation bytes are expected, and what range of values those continuation bytes may have. Anything outside these ranges is considered invalid and terminates the algorithm. See Algorithm~\ref{algo:branchvalidation}.


Such a relatively simple algorithm is commonly found inside popular software. As a reference,  we use the validation function from the Fuchsia operation system~\cite{singh2019fuchsia} by Google. The Fuchsia engineers have benchmarked this function with some care. It follows closely our description.

\begin{algorithm}[htb]
\begin{algorithmic}
\For {each byte $b$ in the UTF-8 sequence}
    \Switch{$b$}
        \CCase{\utfzerobrange{}{00000000}{01111111}} Continue the loop. \EndCCase \Comment{ASCII \codepointrange{0}{7F}}
        \Case{\utfzerobrange{110}{00010}{11111}} \Comment{2-Byte \codepointrange{80}{7FF}}
            \State Load the next byte $c_1$, returning an error if it is not a continuation (\utfzerob{10}{}).
        \EndCase
        \Case{\utfzerob{1110}{0000}} \Comment{3-Byte Low \codepointrange{800}{FFF}}
            \State Load the next two bytes $c_1$ and $c_2$, returning error on EOF or if not continuations (\utfzerob{10}{}).
            \IIf{$c_1 \in$ \utfzerobrange{10}{000000}{011111}} Return error (Overlong). \EndIIf
        \EndCase
        \Case{\utfzerob{1110}{1101}} \Comment{3-Byte \codepointrange{D000}{D7FF}}
            \State Load the next two bytes $c_1$ and $c_2$, returning error on EOF or if not continuations (\utfzerob{10}{}).
            \IIf{$c_1 \in$ \utfzerobrange{10}{100000}{111111}} Return error (Surrogate). \EndIIf
        \EndCase
        \Case{\utfzerobrange{1110}{0001}{1100} or \utfzerobrange{1110}{1110}{1111}} \Comment{3-Byte \codepointrange{1000}{CFFF}, \codepointrange{E000}{FFFF}}
            \State Load the next two bytes $c_1$ and $c_2$, returning error on EOF or if  not continuations (\utfzerob{10}{}).
        \EndCase
        \Case{\utfzerob{11110}{000}} \Comment{4-Byte \codepointrange{10000}{3FFFF}}
            \State Load the next three bytes $c_1$, $c_2$ and $c_3$, returning error on EOF or if not continuations (\utfzerob{10}{}).
            \IIf{$c_1 \in$ \utfzerobrange{10}{000000}{001111}} Return error (Overlong). \EndIIf
        \EndCase
        \Case{\utfzerobrange{11110}{001}{011}} \Comment{4-Byte \codepointrange{40000}{FFFFF}}
            \State Load the next three bytes $c_1$, $c_2$ and $c_3$, returning error on EOF or if not continuations (\utfzerob{10}{}).
        \EndCase
        \Case{\utfzerob{11110}{100}} \Comment{4-Byte \codepointrange{100000}{10FFFF}}
            \State Load the next three bytes $c_1$, $c_2$ and $c_3$, returning error on EOF or if not continuations (\utfzerob{10}{}).
            \IIf{$c_1 \in$ \utfzerobrange{10}{100000}{111111}} Return error (Too Large). \EndIIf
        \EndCase
        \EElse {Return error} \EndEElse \Comment{Too Long, 5+ Byte}
    \EndSwitch
\EndFor
\State Return that the sequence is valid.
\end{algorithmic}
\caption{Branchy Range Validator algorithm.  Byte values are treated as integers in $[0,255]$.}
\label{algo:branchvalidation}
\end{algorithm}

\paragraph{ASCII Optimization}

In many practical instances, UTF-8 contains long strings of ASCII, where a vectorized ASCII check can save us many loop iterations. If there are at least eight bytes to read, we load them into an 8~byte register and quickly check whether any of the characters are non-ASCII (i.e., if they have header bits). This can be done with a simple AND operation against the 8~byte integer value \texttt{0x8080808080808080}, followed by a comparison with zero. If we have found eight consecutive ASCII characters, we just advance the byte pointer by eight and resume the loop, checking again for the presence of eight more bytes.
We find in practice that it is better to go even wider: we check that the next 16~bytes are ASCII by loading the next 16~bytes into two 8-byte registers, computing their bitwise OR and then using the same 8-byte mask \texttt{0x8080808080808080}. We refer to this algorithm, with 16-byte ASCII test, as branchy-ascii. Though we could further widen this approach, we observe poorer performance with wider ASCII checks (i.e., 32~bytes) on realistic data.

\section{Finite-State Machine}
\label{sec:finitestate}

Even on completely valid input, the Branchy Range Validator branches based on the width of each character. This can cause processor stalls when character widths vary (a frequent occurrence in non-ASCII text). To eliminate this issue, we consider a  finite-state machine-based approach.

We could not find an existing finite-state UTF-8 validator. We adapt a UTF-8 decoder proposed by Hoehrmann~\cite{Hoehrmann}. This state machine can be in one of nine possible states:

\begin{itemize}
    \item State \ValidState{} indicates the file is valid to this point. We always begin with \ValidState{}.
    \item States ``1~more'', ``2~more'' and ``3~more'' indicate the number of remaining bytes in the character, and that they can be any value.
    \item Range error states ``3-Byte Overlong'' and ``3-Byte Surrogate'', ``4-Byte Overlong'', and ``4-Byte Too Large'' indicate that there are 2~or 3~bytes remaining in the character, but that the next byte must be checked against a specific range to ensure we do not accept certain invalid values (i.e., it must be a continuation byte, but cannot be just \emph{any} continuation byte).
    \item The \ErrorState{} state indicates we have detected an error. Once it reaches this state, it never leaves.
\end{itemize}
Table~\ref{tab:state-transitions} describes the transitions. As previously stated, the \ErrorState{} state is ``sticky'', with any byte leading back to \ErrorState{}.
When the state is \ValidState{}, the byte is treated as the first byte of a character: it is possible for the state to transition to  any of the nine possible states.
From states ``3~more'', ``4-Byte Overlong'', and ``4-Byte Too Large'', we always either transition to an error or to the state ``2~more''.
When the state is ``2~more'', ``3-Byte Overlong'' or ``3-Byte Surrogate'', we always either transition to an error or to the state ``1~more''.
When the state is ``1~more'', we always transition to an error or to the state \ValidState{}

To quickly compute the transition, we need to classify any new byte into one of
these categories:
\begin{enumerate}
    \item Continuation Low (\utfbinaryrange{10}{000000}{011111}), 
    \item Continuation (\utfbinaryrange{10}{010000}{001111}), 
    \item Continuation High (\utfbinaryrange{10}{100000}{111111}),  
\end{enumerate}
and each of the nine categories corresponding to the last column of Table~\ref{tab:state-transitions} (ASCII,\utfbinaryrange{110}{00010}{11111}, etc. ).
Thus only twelve categories in total are needed. To map any of the 256~possible byte values to
one of these twelve categories without branching, we use a 256-entry lookup table.
We combine efficiently the resulting category (e.g, as an integer between 0 to 11) with 
the state (e.g., as a multiple of 12, from 0 to 108) with an addition, so
that state + class is always a distinct value. Finally, the combined value is used to look up the next state in another table.

\begin{table}\centering
    \caption{Finite-State Machine Transitions. We distinguish between three types of continuation bytes:  Continuation Low (\utfbinaryrange{10}{000000}{001111}), Continuation (\utfbinaryrange{10}{010000}{001111}), and Continuation High (\utfbinaryrange{10}{100000}{111111}). When in \ValidState{} and encountering a non-continuation byte, we determine the next state by using the last column (1st Byte).}
    \label{tab:state-transitions}
\begin{tabular}{l|llllp{2.5cm}}
\toprule
State & \parbox[t]{1.3cm}{Leading Byte}&  \parbox[t]{1.3cm}{Continuation Low}& \parbox[t]{1.3cm}{Continuation} & \parbox[t]{1.3cm}{Continuation High} & 1st Byte \\
\midrule
\ValidState{} & \parbox[t]{1.3cm}{\linespread{0.5}\normalsize\raggedright{1st Byte}\vspace{0.5em}} & \ErrorState{}  & \ErrorState{} & \ErrorState{} & \texttt{00000000}\ldots\texttt{01111111} \\
1~more & \ErrorState{}  & \ValidState{} & \ValidState{} & \ValidState{} & \utfbinaryrange{110}{00010}{11111} \\
2~more & \ErrorState{} & 1~more & 1~more & 1~more & \parbox[t]{4cm}{\normalsize\raggedright\utfbinaryrange{1110}{0001}{1100} \utfbinaryrange{1110}{1110}{1111}} \\
3~more & \ErrorState{}& 2~more & 2~more & 2~more & \utfbinaryrange{11110}{001}{011} \\
3-Byte Overlong &  \ErrorState{} & \ErrorState{} & \ErrorState{}  & 1~more & \utfbinary{1110}{0000} \\
3-Byte Surrogate & \ErrorState{}& 1~more & 1~more & \ErrorState{} & \utfbinary{1110}{1101} \\
4-Byte Overlong & \ErrorState{} & \ErrorState{} & 2~more & 2~more & \utfbinary{11110}{000} \\
4-Byte Too Large & \ErrorState{}& 2~more & \ErrorState{} & \ErrorState{} & \utfbinary{11110}{100} \\
\ErrorState{} & \ErrorState{} & \ErrorState{}& \ErrorState{} & \ErrorState{} & \parbox[t]{4cm}{\normalsize\raggedright\utfbinaryrange{110}{00000}{00001} \utfbinaryrange{1111}{0101}{1111}} \\
\bottomrule
\end{tabular}
\end{table}

Each byte processed requires two  memory loads from small tables: one to categorize the byte and one to determine the new state. The first lookup in the 256-entry table only depends on the character value and may begin before we have the new state.\footnote{Current commodity processors can have several memory requests in flight at the same time.} However, there is a critical data dependency between the successive table lookup that update the state.

We could also combine the two small tables into a single large one (with $9 \times 256$~entries) to halve the number of memory loads: in our tests, it is no faster and uses more memory. This lack of benefit is expected since we do not remove the critical data dependency tied to state updates.

Such  table-based algorithms are crucially dependent on the latency of loads: at least three cycles on x64 processors. To compensate, when the input string is sufficiently long (\SI{32}{byte}), we divide the strings into three distinct regions of nearly equal size, all of them starting with a leading byte. We then run three interleaved versions of the algorithm, loading three distinct bytes from the three regions, and updating three distinct states. We arrived at the number three experimentally, by trying the different variants (1, 2, 3, 4, \ldots interleaved versions).
We call the resulting algorithm \emph{finite-state}.

It would be possible to add branching to finite-state to accelerate ASCII decoding. However, we would then lose the core conceptual benefit  of the finite-state approach: the lack of branches.

\section{The \Lookup{} Algorithm}

The \Lookup{} algorithm mitigates the finite-state machine's memory latency using small lookup tables that fit in SIMD registers. It also vectorizes the problem, validating many bytes of input at a time.

We rely on a key property of the validation problem: nearly all invalid UTF-8 cases can be detected by looking at the first two~bytes of a character (in fact, the first 12~bits---see Table~\ref{tab:invalid2bytepatterns}). The only cases that cannot be detected in 2~bytes are sequences with extra or missing third, fourth or fifth bytes. All \emph{those} can be detected with 4~bytes (see Table~\ref{tab:invalid34bytepatterns}).

SIMD registers on a given architecture might span $w=16$~bytes (e.g., ARM NEON, Intel SSE2), $w=32$~bytes (e.g., AVX/AVX2) or even wider (e.g., AVX-512), allowing the algorithm to check more bytes at once, but the width is irrelevant for algorithmic purposes.

We load the file $w$~bytes at a time into SIMD register $v_1$. The previous input is kept in register $v_0$. On the first iteration, $v_0$ is filled with zero (the ASCII null character). 

Instead of branching on a error conditions, we use an ``error register'' that is  non-zero if and only if an error is detected. The error register is similar to the state variable in the finite-state algorithm (\S~\ref{sec:finitestate}). To modify the
error register,  we use a bitwise OR between an expression that is non-zero if and only if an error is detected. In this manner, we avoid branches. A single check at the end can determine whether there was an error. 

\begin{table}\centering
    \caption{Invalid 1--2~byte UTF-8 Sequences.}
    \label{tab:invalid2bytepatterns}
    \begin{tabular}{lll}
        Error & \multicolumn{2}{l}{UTF-8} \\
        \midrule
        Overlong (2--Byte) & \utfbinaryrange{110}{00000}{00001} & \\
        Overlong (3--Byte) & \utfbinary{1110}{0000} & \utfbinary{10}{0} \\
        Overlong (4--Byte) & \utfbinary{11110}{000} & \utfbinary{10}{0} \\
        Too Short (Missing 2nd Byte) & \utfbinary{11}{} & \utfbinary{0}{} \\
        Too Long (ASCII + Continuation) & \utfbinary{0}{} & \utfbinary{10}{} \\
        Surrogate & \utfbinary{1110}{1101} & \utfbinary{10}{1} \\
        Too Large & \utfbinary{11110}{100} & \utfbinary{10}{1}\\
        Too Large & \utfbinaryrange{11110}{101}{111} & \\
        Too Large (5+--Byte) & \utfbinary{111111}{} & \\
    \end{tabular}
\end{table}
\begin{table}\centering
    \caption{Invalid 3--4~byte UTF-8 Patterns.}
    \label{tab:invalid34bytepatterns}
    \begin{tabular}{lllll}
        Error & \multicolumn{4}{l}{UTF-8} \\
        \midrule
        Too Long (Extra 3rd Byte) & \utfbinary{11}{} & \utfbinary{10}{} & \utfbinary{10}{} & \\
        Too Long (Extra 4th Byte) & \utfbinary{111}{} & \utfbinary{10}{} & \utfbinary{10}{} & \utfbinary{10}{} \\
        Too Long (Extra 5th Byte) & \utfbinary{10}{} & \utfbinary{10}{} & \utfbinary{10}{} & \utfbinary{10}{} \\
        Too Short (Missing 3rd Byte) & \utfbinary{111}{} & \utfbinary{10}{} & \utfbinary{0}{} & \\
        Too Short (Missing 3rd Byte) & \utfbinary{111}{} & \utfbinary{10}{} & \utfbinary{11}{} & \\
        Too Short (Missing 4th Byte) & \utfbinary{11110}{} & \utfbinary{10}{} & \utfbinary{10}{} & \utfbinary{0}{} \\
        Too Short (Missing 4th Byte) & \utfbinary{11110}{} & \utfbinary{10}{} & \utfbinary{10}{} & \utfbinary{11}{} \\
    \end{tabular}
\end{table}

\subsection{Invalid 2--Byte Sequences}
\label{sec:invalidtwo}
After loading $v_1$, we detect all invalid 2-byte sequences at once using vectorized classification, a concept we documented in earlier work~\cite{langdale2019parsing}. In this scheme, we classify several values at once by doing combining vectorized table lookups. Compared to earlier work, this particular application of vectorized classification uses three different table lookups, instead of merely two. Both ARM and x64 systems have vectorized lookup tables allowing us to use a 4-bit value (nibble) stored in byte as an index into a 16-byte register (e.g., \texttt{vpshufb} in AVX2). Even when the source register has 16~or 32~bytes, the 16~or 32~lookups can occur at once, using a single instruction.

For UTF-8, we create three 16-entry lookup tables that map to 8-bit values. Bits 0--6 of these values, when set, indicate a partial match against one of seven error patterns. These patterns were chosen to encompass all possible 2-byte errors (Table~\ref{tab:errorpatterns}). The high and low nibbles of each byte, as well as the low nibble of the next byte, are looked up in their respective tables. To get the low nibble of a byte, we mask an existing register (AND 0xF). To get the high nibble of each byte in a register, we shift its bytes right by 4 bits.\footnote{Under x64, we lack a byte-wise vectorized shift but we can shift 16-bit words with a vector instruction (e.g.,  \texttt{vpsrlw}) and apply a mask to select the low nibble. }

If a bit in the range 0--6 is set in all three looked-up patterns for a byte as checked with the AND instruction,\footnote{We use two vector AND instructions to combine the three patterns, but for processors supporting it,  a single  AVX-512 instruction (\texttt{vpternlog}) would suffice.} that byte (and the UTF-8) is considered invalid. Bit~7 is used to identify a pair of continuation bytes, which is used in \S~\ref{sec:threefour} to evaluate long invalid 3--4~byte sequences, but by itself, bit~7 is not considered an error.

There will always be a pair of bytes straddling the two SIMD registers, which need to be validated as well. To get the correct first byte to match against each second byte, we shift the input one byte to the right, ``shifting in'' the last byte of the previous input as we do so. Under ARM NEON, we use the \texttt{ext} instruction. Under x64, a single instruction in the 128-bit case (\texttt{palignr}) or two in the 256-bit case (\texttt{vpalignr} and \texttt{vperm2i128}) suffice.

\begin{table}\centering
    \caption{List of 2-Byte error patterns. Any pair of bytes matching one of these patterns is considered invalid except for the last row (bit~7).}
    \label{tab:errorpatterns}
    \begin{tabular}{rrlll}
        Error Bit & Error & \multicolumn{1}{l}{Byte 1} & Byte 2 \\
        \midrule
        0 & Too Long (ASCII + Continuation) & \utfbinary{0}{} & \utfbinary{10}{} \\
        1 & Too Short (Missing 2nd Byte) & \utfbinary{11}{} & \utfbinary{0}{} \\
         & \emph{+} & \utfbinary{11}{} & \utfbinary{11}{} \\
        2 & Overlong (2--Byte) & \utfbinaryrange{110}{00000}{00001} & \utfbinary{10}{} \\
        3 & Surrogate & \utfbinary{1110}{1101} & \utfbinary{10}{1} \\
        4 & Overlong (3--Byte) & \utfbinary{1110}{0000} & \utfbinary{10}{0} \\
        5 & Overlong (4--Byte) & \utfbinary{11110}{000} & \utfbinary{10}{00} \\
        & \emph{+ Too Large} & \utfbinaryrange{1111}{0101}{1111} & \utfbinary{10}{00} \\
        6 & Too Large & \utfbinaryrange{1111}{0100}{1111} & \utfbinaryrange{10}{01}{11} \\
        7 & Two Continuations (Not An Error) & \utfbinary{10}{} & \utfbinary{10}{} \\
    \end{tabular}
\end{table}

The original Table~\ref{tab:invalid2bytepatterns} contains nine error patterns. We consolidated these into seven error patterns that cover all possible errors, so that we could save a bit for continuation pairs. The problem of finding a minimal cover is NP-hard~\cite{karp1972reducibility}, but is thankfully inexpensive with only seven patterns.
Fig.~\ref{fig:pseudocpp} illustrates our routine using pseudocode. It closely matches our C++ code. Table~\ref{tab:illustraiton} provides a processing example, with the corresponding variable names, starting with the null-terminated string ``9\cent{}\begin{CJK}{UTF8}{bsmi}鏡\end{CJK}\Smiley[][yellow]''. Because we assume that it begins the stream, we set the previous input to zeroes.
We compute three vectors made of nibbles and from these three vectors we derive three lookup results (byte\_1\_high,  byte\_1\_low, byte\_2\_high).
The
final result is the bitwise AND of the three lookup results. It is made entirely of zeroes except at three locations
corresponding to the last byte of the character \begin{CJK}{UTF8}{bsmi}鏡\end{CJK} and
to the last two bytes of the character \Smiley[][yellow].

\begin{figure}[b]
\lstset{escapechar=@,style=customc}
\begin{lstlisting}
  simd8<uint8_t> classify(simd8<uint8_t> input, simd8<uint8_t> previous_input) {
    // shift the input by 1 byte, shifting in the last byte of the previous input
    auto prev1 = input.prev<1>(previous_input);
    auto byte_1_high = prev1.shift_right<4>().lookup_16(table1);
    auto byte_1_low = (prev1 & 0x0F).lookup_16(table2);
    auto byte_2_high = input.shift_right<4>().lookup_16(table3);
    return (byte_1_high & byte_1_low & byte_2_high);
  }
\end{lstlisting}
\caption{Pseudocode corresponding to the vectorized classification routine\label{fig:pseudocpp}}
\end{figure}

\begin{table}\centering
    \caption{Vectorized classification example using the notation of Fig.~1 for the null-terminated string ``9\cent{}\begin{CJK}{UTF8}{bsmi}鏡\end{CJK}\Smiley[][yellow]''. We use hexadecimal byte values.}
    \label{tab:illustraiton}
    \begin{tabular}{llllllllllll}
    &\multicolumn{1}{l}{'9'} & \multicolumn{2}{l}{'\cent{}'} & \multicolumn{3}{l}{'\begin{CJK}{UTF8}{bsmi}鏡\end{CJK}'} & \multicolumn{4}{l}{'\Smiley[][yellow]'} & 0\\\midrule
input &     39 & C3 & A7 & E9 & 8F & A1 & F0 & 9F & 98 & 80 & 00 \\
previous\_input (set to zero)  &    00 & 00& 00& 00& 00&00& 00& 00& 00& 00& 00 \\
prev1 (shifted input) &    00& 39& C3& A7& E9&8F& A1& F0& 9F& 98& 80 \\
high nibbles: prev1.shift\_right<4>()  &   00& 03& 0C& 0A& 0E& 08& 0A& 0F& 09& 09& 08 \\
low nibbles:  (prev1 \& 0x0F)  &  00& 09& 03& 07& 09& 0F& 01& 00& 0F& 08& 00\\
high nibbles: input.shift\_right<4>()  &  03& 0C& 0A& 0E& 08& 0A& 0F& 09& 09& 08& 00\\
lookup result: byte\_1\_high  &   02& 02& 21& 80& 15& 80& 80& 49& 80& 80& 80 \\
lookup result: byte\_1\_low  &  E7& CB& 83& CB& CB& CB& A3& E7& CB& CB& E7\\
lookup result: byte\_2\_high &  01& 01& BA& 01& E6& BA& 01& AE& ae& E6 &01\\
(byte\_1\_high \& byte\_1\_low \& byte\_2\_high)  &  00& 00& 00& 00& 00& 80& 00& 00& 80& 80& 00\\
    \end{tabular}
\end{table}


 
\subsection{Invalid 3--4~Byte Sequences}
\label{sec:threefour}
All remaining checks are invalid 3--4~byte sequences, which either have too many continuations, or not enough (Table~\ref{tab:invalid34bytepatterns}).
We first get a list of byte indexes where we expect to find two continuations in a row, which can only be found in 3--4~byte sequences. We can compute these indexes with a pair of shifts and comparisons: we expect two continuations if the previous byte matches \utfbinary{111}{}, or if the byte before that matches \utfbinary{1111}{}.
We then compare these indexes to the locations where we have two consecutive continuations, as detected by bit~7 from our vectorized classification (see  \S~\ref{sec:invalidtwo}). If these two lists differ in any respect, the UTF-8 is invalid. 

Under x64, we lack unsigned comparison instructions, which are needed to check the \utfbinary{111}{} and \utfbinary{1111}{} ranges. However, we can emulate them in various efficient ways. For example, to compute the equivalent of $m_3(v_{i-1}, v_i)\geq 0xF0$, we can use the saturated subtraction of $m_3(v_{i-1}, v_i)$ with $0xF0 - 1$ which results in a number greater than~0 where and only where $m_3(v_{i-1}, v_i)\geq 0xF0$. Thus we can compute two saturated subtraction, combine the two results using one bitwise OR\@. We are then left to apply a mask to set just the most significant bit of each byte where a continuation byte  should appear.

 \subsection{Incomplete Stream}
 \label{sec:incompletestream}

At the end of the stream, we may not have enough bytes to fill an entire SIMD register. If that is the case, we may simply virtually fill the leftover bytes with any ASCII character (such as zero). But even when we have enough bytes to fill a whole register, we still have to check that the data stream does not terminate with an incomplete code point. Furthermore, the 2-byte check (\S~\ref{sec:invalidtwo}) may allow a byte value larger than the maximum (0xF4) if it occurs as the last byte of a stream. Thankfully, it not difficult to guard against both problems. We just have to check that the last byte in the last register is strictly smaller than 0xC0 (using an unsigned comparison), that the second last byte is strictly smaller than  0xE0, and that the third last byte is  strictly smaller than  0xF0. A single vectorized unsigned comparison is sufficient. On x64 processors, there is no unsigned comparison instruction, but we can use an efficient alternative such as an unsigned vectorized maximum instruction followed by a comparison.

 \subsection{ASCII}

Because a lot of content might be ASCII, it sometimes pay to check whether the current register is made of ASCII bytes. We can efficiently check whether a given register is made of all ASCII bytes: we can check that the byte values are all negative (using two's complement). When they are ASCII, we may then use a fast path: we omit the vectorized classification and the check on the continuation bytes.
However, before we do so, we need to check that the previous register did not end with an incomplete code point (\S~\ref{sec:incompletestream}). Doing such checks vector register by vector register might be too expensive. When the location of the ASCII blocks is hard to predict, these checks could create many mispredicted branches. Instead, we group the registers in blocks of 64~bytes.\footnote{A 64-byte block size matches the length of a cache line on most x64 processors.} We check whether the whole block (64~bytes) is ASCII\@. In such a case, we also
need to verify that the preceding block finished with  complete code points (\S~\ref{sec:incompletestream}), and if so we do not need any further checks. To validate a whole block of SIMD registers, we could do one comparison per register and then aggregate the result of the comparisons with a bitwise OR\@. This results in roughly two instructions per register.\footnote{A single  AVX-512 instruction (\texttt{vpternlog}) might also replace two bitwise OR.}
 Instead, it is more advantageous to compute the bitwise OR between all registers and then to do one comparison: the block is all ASCII if and only if the bitwise OR are non-negative.\footnote{Checking that the register is entirely non-negative requires few instructions: e.g.,  a \texttt{pmovmskb}/\texttt{vpmovmskb} instruction under x64  or a \texttt{umaxv} instruction under ARM NEON, followed by a comparison.} This results in nearly half the number of instructions.

When our input is made entirely of either ASCII characters, or of sequences containing non-ASCII characters, the fast ASCII path is either always called or never called. Thus the branches are easily predicted with high accuracy. In other scenarios,  we have to rely on the processors' sophisticated branch predictor for performance.

\section{Experiments}

We wish to benchmark our algorithms on common x64 processors.
Recent Intel processors are often based on the Skylake microarchitecture or similar variations. AMD recently introduced a competitive microarchitecture (Zen~2); we use a server version of this architecture. 
We summarize the characteristics of our hardware platforms in
 Table~\ref{tab:test-cpus}. The Intel processor has a slightly higher frequency, but the AMD processor has a more recent microarchitecture. Both processors have \SI{32}{\kilo\byte} of (data) L1 cache. The AMD processor has more L2 cache (\SI{512}{\kilo\byte} vs. \SI{256}{\kilo\byte}). 
 
\begin{table*}
\caption{\label{tab:test-cpus} Hardware 
}
\centering
\begin{minipage}{\textwidth}
\centering
\begin{tabular}{cccccc}\toprule
Processor   & Base Frequency & Max.   & Microarchitecture                           & Memory  & Compiler\\ \midrule
Intel i7-6700  & 3.4\,GHz   & 3.7\,GHz & Skylake (x64, 2015) & DDR4 (2133\,MT/s) & GCC 9.3   \\
AMD EPYC 7262  (Rome) & 3.2\,GHz  & 3.4\,GHz & Zen~2 (x64, 2019) &  DDR4 (3200\,MT/s)  & GCC  9.3 \\
\bottomrule
\end{tabular}
\end{minipage}
\end{table*}

Our software is  written using C++ (GNU GCC 9.3) and we use Linux Ubuntu (20.10). We compile the code for best performance with the \texttt{-O3} flag. All code is single-threaded and free from disk or network access. We expect all processed inputs to be in CPU cache, by design.  Our software, including benchmarking code and corresponding instructions, is freely available.\footnote{\texttt{https://github.com/lemire/validateutf8-experiments}}

Our benchmarking code is \emph{instrumented}: we use the performance counters of the processors to record the number instructions retired, the number of cycles and the number of mispredicted branches. Performance timings are heavily skewed to the right and they do not follow a normal distribution~\cite{hoefler2015scientific}. Following our earlier work~\cite{langdale2019parsing}, we run each test many times (1000) and compute both the best (smallest) and the average timing. We find that the average and the smallest timing coincide (within \SI{1}{\percent}).

Nevertheless, we  still  slightly overestimate the number of elapsed cycles and the duration of the tests (by about \SIrange{10}{30}{\nano\second}). In effect, we get slightly worse performance numbers. When the tests last a sufficient long time, we can simply 
ignore the effect since the overhead is negligible. However, we cannot ignore this measurement overhead on small inputs (e.g., less than \SI{1}{\kibi\byte}) when using fast functions like \Lookup{}. To compensate, we use the following strategy when necessary: we select a string that is twice as long as the desired size, and then we select a valid substring having nearly (within a few bytes) the desired size. We run both benchmarks and subtract the timings. We report the difference as a compensated measure.

We use the AVX2 instruction set and 256-bit vector registers. The Intel processor is subject to \emph{downclocking}: with AVX2 instructions   using floating-point operations and multiplications, the processor may reduce its frequency temporarily. However, the \Lookup{} algorithm does not use multiplication or floating-point operations, and it therefore does not trigger downclocking. We consistently achieve the maximum frequency of the processors.  We record the effective processor frequency and find it to be constant within a small margin of error ($\approx$\SI{1}{\percent}).

\subsection{Mispredicted Branches}

When benchmarking functions involving branches, we must consider the ability of the processor to \emph{learn branches}. When executing the same function, over the same data, repeatedly, we may expect the processor to eventually learn to predict the branches. This is unlikely to happen if the input is large and irregular, but a poorly constructed benchmark made of small or predictable input can lead to spurious results and conclusions.

\begin{figure*}\centering
\subfloat[AMD Rome  (Zen 2; \SI{3.4}{\giga\hertz})]{%
\includegraphics[width=0.49\textwidth]{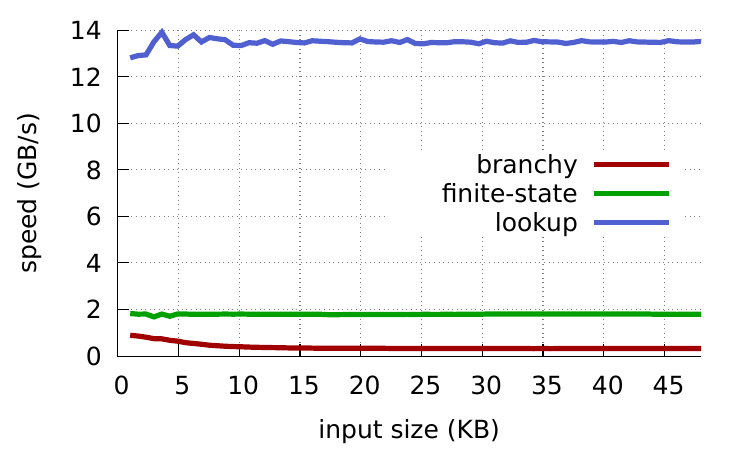}
}\subfloat[Intel Skylake  (\SI{3.7}{\giga\hertz})]{%
\includegraphics[width=0.49\textwidth]{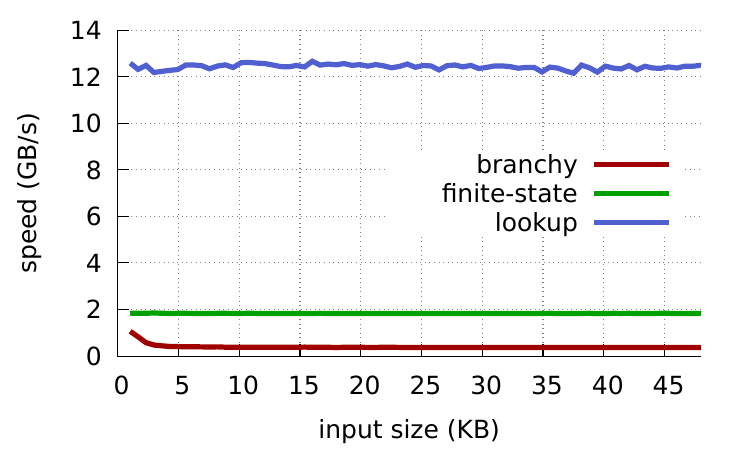}}
\caption{\label{fig:randspeed}Processing speed for random UTF-8 inputs of various lengths (one-and-two-byte code points). }
\end{figure*}

\begin{figure*}\centering
\subfloat[AMD Rome]{%
\includegraphics[width=0.49\textwidth]{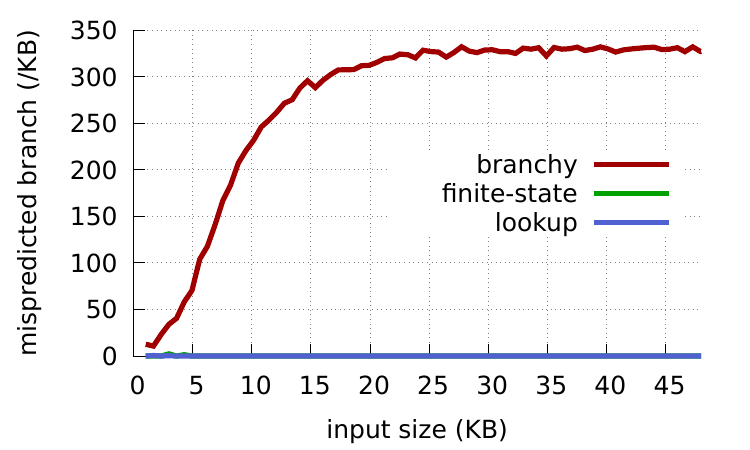}
}\subfloat[Intel Skylake]{%
\includegraphics[width=0.49\textwidth]{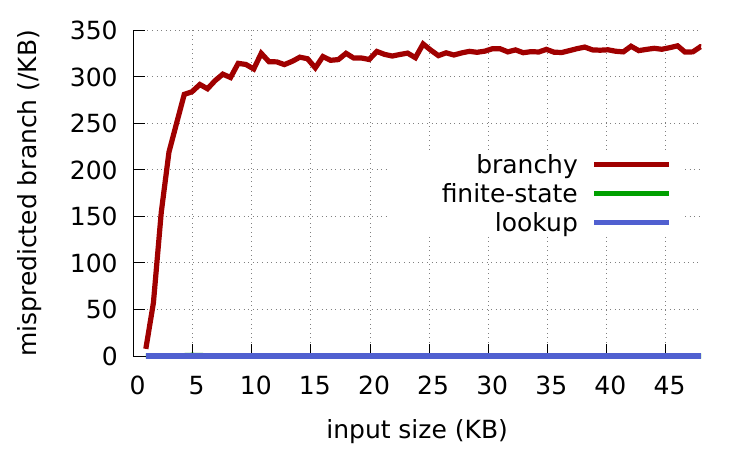}}
\caption{\label{fig:randmis}Number of mispredicted branches per kilobytes for random UTF-8 inputs of various lengths. }
\end{figure*}

We generated random UTF-8 strings of various lengths, using random code points. We pick each successive code point to have either one or two-byte length in UTF-8. Once we have generated a string of a given length, we repeatedly validate it (in a tight loop) while measuring speed and number of mispredicted branches. We pick the run with the best speed for each given length. We find that branchy is faster on small inputs: see Fig~\ref{fig:randspeed}. The reason becomes clear when looking at the number of mispredicted branches (Fig.~\ref{fig:randmis}). Because of how we designed our inputs, we should expect a mispredicted branch every three bytes, so $\approx$333~mispredicted branches per kilobyte.

We find that both the AMD Rome and the Intel Skylake processors have far fewer than 333~mispredicted branches per kilobyte on short inputs, an indication that the branch predictor \emph{has learned} the content of the string. The branch predictors work well up until about \SI{30}{\kilo\byte}. Observe that the AMD Rome processor is better at predicting branches than the Intel Skylake processor. 

\subsection{Realistic Inputs}

Of course, the validation performance depends on the input.
We use two sizeable input files: a JSON file (twitter.json, \SI{617}{\kibi\byte}) produced from the Twitter API and an HTML file (hongkong.html, \SI{1.8}{\mebi\byte}) captured from the corresponding Wikipedia entry. See Table~\ref{tab:files}.

The speed of the branchy and of the finite-state validators are similar on the two test files, at roughly \SI{2}{\gibi\byte\per\second}. These files are an instance where branchy-ascii is advantageous because they contain long sequences of ASCII strings. It is almost twice as fast as branchy. Though the Intel processor has a higher clock speed (by about \SI{10}{\percent}), 
the AMD processor is more than \SI{50}{\percent} faster when running the branchy validator. We also find that the AMD
Rome has fewer mispredicted branches per kilobyte: 3.9 versus 4.6 (Intel) for twitter.json and 8.2 versus 7.6 for hongkong.html.

Under both AMD Rome and Intel Skylake, we find that \Lookup{} retires slightly under 0.4~instructions per byte for both files. Yet
the throughput of \Lookup{} is higher under twitter.json than under hongkong.html. The explanation for this apparent contradiction
lies in the fact that the hongkong.html file triggers many more branch mispredictions.

The number of mispredicted branches per byte is tiny with twitter.json under both processors.
For hongkong.html, we observe 2.0~mispredicted branch per kilobyte on AMD Rome, and slightly more on Intel Skylake (2.6).
We find that the AMD processors is faster than the Intel processor 
when running \Lookup{} (\SIrange{5}{15}{\percent}) despite a lower clock speed.
Under the Intel processor, the \Lookup{} validator comes close to matching the speed of the \texttt{memcpy} function when processing the file twitter.json: \SI{24}{\gibi\byte\per\second} versus \SI{36}{\gibi\byte\per\second}.







\begin{table}\centering
\caption{\label{tab:files}Throughput in \si{\gibi\byte\per\second} to  validate UTF-8 files. The original files are valid UTF-8. We also benchmark the C function \texttt{memcpy}, copying the content to a temporary buffer.}
\subfloat[AMD Rome (Zen 2; \SI{3.4}{\giga\hertz})]{%
\begin{tabular}{lrr}\toprule
validator & twitter.json & hongkong.html \\\midrule
\texttt{memcpy}   & 48 & 48 \\ \midrule
branchy   & 2.5 & 2.3 \\ 
branchy-ascii   & 4.4 & 4.3 \\ 
finite-state & 2.0 & 2.0\\
\Lookup{}  & 28 & 18 \\ \bottomrule
\end{tabular}
}
\hspace{0.1\textwidth}
\subfloat[Intel Skylake (\SI{3.7}{\giga\hertz})]{%
\begin{tabular}{lrr}\toprule
validator & twitter.json & hongkong.html \\\midrule
\texttt{memcpy}   & 36 & 36 \\ \midrule
branchy   & 1.6 & 1.6\\ 
branchy-ascii   & 4.0 & 4.4\\ 
finite-state & 1.8 & 1.8 \\
\Lookup{} & 24 & 17 \\ \bottomrule
\end{tabular}
}
\end{table}

\subsection{Randomized Inputs}

To test our functions with different inputs, it is useful to generate synthetic UTF-8 data.
If we select to generate code-point values spanning 1--3~bytes, we randomly pick, for each code point, a byte length in the range 1--3, uniformly at random. The generator produces bytes by adding new code-point values until we have generated \SI{16}{\kilo\byte}. In general, the final string may exceed \SI{16}{\kilo\byte} by up to 3~bytes. A data input \SI{16}{\kilo\byte} is long enough to prevent the branch predictor from \emph{learning the input}, but short enough to fit in L1~CPU cache.
See Table~\ref{tab:randfast}.

The finite-state approach offers a flat performance of \SI{1.8}{\gibi\byte\per\second} irrespective of the input source. Such data independence is expected given that the algorithm is essentially free from branches. The branchy-ascii approach does well on the ASCII-only inputs (\SIrange{14}{15}{\gibi\byte\per\second}) and roughly as well as branchy on non-ASCII synthetic inputs. The \Lookup{} algorithm dominates, being 30~times faster than branchy and branchy-ascii on non-ASCII inputs, and six~times faster than finite-state.

On ASCII inputs,  the \Lookup{} function is  faster than the \texttt{memcpy} function, achieving \SI{66}{\gibi\byte\per\second} on the AMD processor and \SI{59}{\gibi\byte\per\second} on the Intel processor.
In the special case where we expect our strings to be pure ASCII, we could design even faster functions with and without SIMD instructions but our purpose is UTF-8 validation.

\begin{table}\centering
\caption{\label{tab:randfast}Throughput in \si{\gibi\byte\per\second} to  validate UTF-8 randomized inputs where code-point values have different byte lengths. We also benchmark the C function \texttt{memcpy}, copying the content to a temporary buffer.}
\subfloat[AMD Rome (Zen 2; \SI{3.4}{\giga\hertz})]{%
\begin{tabular}{lrrrr}\toprule
validator & ASCII & 1--2 bytes & 1--3 bytes & 1--4 bytes\\\midrule
\texttt{memcpy}   & 53 & 53 & 53& 53 \\ \midrule
branchy   & 1.7 & 0.41 & 0.39 & 0.60  \\ 
branchy-ascii   & 14 & 0.33 & 0.42 & 0.63 \\ 
finite-state & 1.8 & 1.8 & 1.8 & 1.8 \\
\Lookup{}  & 66 & 13 &13 & 13   \\ \bottomrule
\end{tabular}
}
\hspace{0.1\textwidth}
\subfloat[Intel Skylake (\SI{3.7}{\giga\hertz})]{%
\begin{tabular}{lrrrr}\toprule
validator & ASCII & 1--2 bytes & 1--3 bytes & 1--4 bytes\\\midrule
\texttt{memcpy}   & 39 & 39 & 39 & 39 \\ \midrule
branchy   & 1.8 & 0.36 & 0.35 & 0.40\\ 
branchy-ascii   & 15 & 0.30 & 0.30 & 0.43 \\ 
finite-state & 1.8 & 1.8 & 1.8 & 1.8 \\
\Lookup{} & 59 & 12 & 12 & 12 \\ \bottomrule
\end{tabular}
}
\end{table}

In Table~\ref{tab:ins}, we present the number of retired instructions per byte. The retired instructions are counted by the processor by excluding speculative execution. That is, the instructions part of a mispredicted branch are not counted. The processors count some fused instructions such as the comparison and jump of a branch as two instructions. It is therefore possible for some code executing tight loop with branches to have high numbers of instructions retired, if these branches are correctly predicted with high probability. 

The AMD Rome and Intel Skylake processors have similar instructions counts, so we only present the numbers for the AMD Rome processors. The reason for the good performance of the \Lookup{} algorithm is clear: it requires far fewer instructions than the alternatives (often ten times fewer). 
In all our tests, irrespective of the input, \Lookup{} requires fewer than one retired instruction per byte.

The number of retired per cycle (Table~\ref{tab:inspercycle}) differs between the two processors with an advantage for the AMD Rome processor with branchy, branchy-ascii and \Lookup{}. Except for the pure ASCII inputs, the \Lookup{} function achieves a high 3.6~instructions per cycle on AMD Rome. In all cases, the   \Lookup{} function benefits from a relatively high number of instructions per cycle (at least 3). 

We also find that the finite-state function has a consistently high number of instructions retired per cycle (3.5). However, it suffers from a high number of instructions per byte (7).

\begin{table}\centering
\caption{\label{tab:ins}Instruction per byte to  validate UTF-8 randomized inputs where code-point values have different byte lengths. As a reference we use the AMD Rome processor.}
\begin{tabular}{lrrrr}\toprule
validator & ASCII & 1--2 bytes & 1--3 bytes & 1--4 bytes\\\midrule \
branchy   & 6.0 & 11 & 12 &12 \\ 
branchy-ascii   & 0.75 & 16 & 17 & 16 \\ 
finite-state & 7.0 & 7.0 & 7.0 & 7.0 \\
\Lookup{}  & 0.21 & 0.97 &0.97 & 0.97  \\ \bottomrule
\end{tabular}
\end{table}

\begin{table}\centering
\caption{\label{tab:inspercycle}Instructions per cycle to  validate UTF-8 randomized inputs where code-point values have different byte lengths. }
\subfloat[AMD Rome (Zen 2; \SI{3.4}{\giga\hertz})]{%
\begin{tabular}{lrrrr}\toprule
validator & ASCII & 1--2 bytes & 1--3 bytes & 1--4 bytes\\\midrule
branchy   & 3.0 & 1.3 & 1.4 & 2.2  \\ 
branchy-ascii   & 4.7 & 1.4 & 1.8 & 2.7 \\ 
finite-state & 3.5 & 3.5 & 3.5 & 3.5 \\
\Lookup{}  & 3.2 & 3.6 &3.6& 3.6   \\ \bottomrule
\end{tabular}
}
\hspace{0.1\textwidth}
\subfloat[Intel Skylake (\SI{3.7}{\giga\hertz})]{%
\begin{tabular}{lrrrr}\toprule
validator & ASCII & 1--2 bytes & 1--3 bytes & 1--4 bytes\\\midrule
branchy   & 3.0 & 1.0 & 1.1 & 1.3\\ 
branchy-ascii   & 4.7 & 1.2 & 1.2 & 1.7\\ 
finite-state & 3.5 & 3.5 & 3.5 & 3.5 \\
\Lookup{} &3.0 & 3.1 &3.1& 3.1 \\ \bottomrule
\end{tabular}
}
\end{table}

\section{Related Work}

There has been much work on the acceleration of text content  using SIMD instructions (e.g., base64~\cite{mula2018faster,mula2020base64}, JSON~\cite{langdale2019parsing}, XML~\cite{Cameron:2008:HPX:1463788.1463811}, HTML~\cite{newMytkowicz:2014:DFM:2541940.2541988}, CVS~\cite{Muhlbauer:2013:ILM:2556549.2556555}). We are not aware of any published work directly related to Unicode validation using SIMD instructions other than our own~\cite{langdale2019parsing}. Cameron~\cite{cameron2008case} has worked on the related problem of UTF-8 to UTF-16 transcoding using SIMD instruction, but their approach is not applicable to high-speed validation. 
There has been some research on the parallelisation of finite-state machines~\cite{newMytkowicz:2014:DFM:2541940.2541988,jiang2017combining} which could be applied to UTF-8 validation.

\section{Conclusion}

The relatively simple algorithm (\Lookup{}) can be several times faster than conventional algorithms at a common task using nothing more than the instructions available on commodity processors. It requires fewer than an instruction per input byte in the worst case. This new algorithm has been adopted by the simdjson library  with good results.\footnote{\url{https://simdjson.org}} A SIMD-based approach like \Lookup{} is especially advantageous in a context where the data is loaded in vector registers in any case---as happens in simdjson.

Intel has produced a new family of instruction sets with wider vector registers and more powerful instructions (AVX-512). 
Future research should assess the benefits of AVX-512 instructions to the problem of UTF-8 validation. In principle, we could expect the performance to double~\cite{mula2020base64}.
Similarly, commodity ARM processors may soon benefit from more powerful instructions and wider registers (e.g., SVE and SVE2)~\cite{stephens2017arm,POHL2020102106}.

\section*{Acknowledgements}

The authors would like to thank T.~Downs for discussions on finite-state validation. 
Our work is inspired by an early high-speed SIMD validator written by K.~Willets that
served as a proof-of-principle. 
We thank Z.~Wegner for demonstrating that our early designs were suboptimal and for
proposing inspiring new designs.
Part of our benchmarking code is derived from code by W.~Mu{\l}a.



\bibliography{utf8_lookup_validation}



\end{document}